\documentstyle[12pt]{article}

\def\o{\over}
\def\b{\begin{equation}}
\def\e{\end{equation}}
\def\l{\label}
\def\kpnn{$K^+\rightarrow\pi^+\nu\bar\nu$\ }
\def\kpn{K^+\rightarrow\pi^+\nu\bar\nu}

\def\klmm{$K_L\rightarrow\mu^+\mu^-$\ }
\def\klm{K_L\rightarrow\mu^+\mu^-}

\begin{document}
\thispagestyle{empty}
\begin{flushright}
 MPI-Ph/93-44 \\
 TUM-T31-44/93 \\
 July 1993
\end{flushright}
\vskip1truecm
\centerline{\Large\bf The Rare Decays \kpnn and \klmm}
\centerline{\Large\bf Beyond Leading Logarithms\footnote[1]{\noindent
   Supported by the German
   Bundesministerium f\"ur Forschung und Technologie under contract
   06 TM 732 and by the CEC science project SC1--CT91--0729.}}
\vskip1truecm
\centerline{\sc Gerhard Buchalla {\rm and} Andrzej J. Buras}
\bigskip
\centerline{\sl Max-Planck-Institut f\"ur Physik}
\centerline{\sl  -- Werner-Heisenberg-Institut --}
\centerline{\sl    P.O.Box 40 12 12, Munich, Germany}
\vskip0.6truecm
\centerline{\sl Technische Universit\"at M\"unchen, Physik Department}
\centerline{\sl D-85748 Garching, Germany}

\vskip1truecm
\centerline{\bf Abstract}
We analyze the branching ratio for the FCNC mode \kpnn in the
standard model with QCD effects taken into account consistently to
next-to-leading order. This involves a two-loop renormalization group
analysis for the charm contribution, presented in this paper, and the
calculation of $O(\alpha_s)$ corrections
to all orders in $m_t/M_W$ for the
top-quark case that we have described elsewhere. The inclusion of
next-to-leading corrections reduces considerably the theoretical
uncertainty due to the choice of the renormalization scales, inherent
in any calculation to finite order in perturbation theory. For \kpnn
this point has not been discussed previously. In particular, the related
uncertainty in the determination of $|V_{td}|$ from $B(\kpn)$ is
reduced from $\sim 30\%$ to $\sim 7\%$ for typical values of the
parameters. Simultaneously also the dependence of $B(\kpn)$ on the choice
of $m_c$ is considerably reduced.
We also give the next-to-leading order expression for the
short-distance part of $\klm$.
Impacts of our calculations on the determination of the unitarity
triangle are presented.
\vfill
\newpage

\pagenumbering{arabic}

\section{Introduction}
The rare decay mode \kpnn is one of the most interesting and promising
processes in the rich field of rare decay phenomenology
\cite{BH,LV} and has already
received considerable attention in the past \cite{IL,EH,DDG,BBH,D,HR}.
The main reasons for
this interest may be summarized as follows:\hfill\break
\kpnn is a "clean" process from the theoretical point of view:
\begin{itemize}
\item
It is short-distance (SD) dominated. Possible long-distance contributions
have been considered in \cite{RS,HL} and found to be negligibly small
(three orders of magnitude smaller than
the SD contribution at the level of branching ratios).
\item
Being a semi-leptonic process, the relevant hadronic operator is just a
current operator whose matrix element can be extracted from the leading
decay $K^+\rightarrow\pi^o e^+\nu$.
\end{itemize}
Consequently this decay can be reliably treated by the available
field theoretical methods.\hfill\break
The effective low-energy interaction mediating \kpnn is a flavor-changing
neutral current (FCNC). While being forbidden at the tree level, FCNC's
can be induced through loops containing virtual heavy quarks. Therefore
\kpnn
\begin{itemize}
\item
probes the standard model at the quantum level, thereby allowing for an
indirect test of high energy scales through a low energy process,
\item
is sensitive to the unknown top quark mass $m_t$ and
\item
depends on the top quark couplings $V_{ts}$ and $V_{td}$.
\end{itemize}
Thus combined theoretical and experimental efforts may yield very
desirable information on parameters of the symmetry breaking sector in
the standard model -- the part of the theory least understood theoretically
and containing the least known parameters such as the top mass and its
CKM couplings. In particular it should be stressed that \kpnn is
probably the cleanest process to search for $V_{td}$.\hfill\break
The \kpnn mode has not yet been observed experimentally. The published
upper bound on the branching ratio is \cite{PDG}
\b\l{exp1} B(\kpn )_{exp}\leq 3.4\cdot 10^{-8} \e
A preliminary result from E787 at Brookhaven \cite{A} gives an improved
value of
\b\l{exp2} B(\kpn )_{exp}\leq 5\cdot 10^{-9} \e
which is still well above the theoretical expectation of $O(10^{-10})$,
but a considerable improvement is expected from upgrades planned
for the coming years and designed to reach the $10^{-10}$ level \cite{K}.
\hfill\break
After these general introductory remarks let us now turn to a more
detailed
discussion of the basic structure of the physics underlying \kpnn.
The low energy effective hamiltonian relevant for the process \kpnn may be
written as follows
\b\l{hxxi} {\cal H}_{eff}={G_F \o{\sqrt 2}}{\alpha\o 2\pi \sin^2\Theta_W}
 \sum_{i=c,t} V^{\ast}_{is}V_{id}\ X(x_i)\
 (\bar sd)_{V-A}(\bar\nu\nu)_{V-A}  \e
where $x_i=m^2_i/M^2_W$ and $X(x)$ is a
monotonically increasing function of $x$ to be given below.
This hamiltonian consists of two parts: a top-quark and a charm-quark
contribution, which originate to lowest order in the standard model from
the Z-penguin- and box-diagrams shown in fig.1. By writing the
hamiltonian in this form, the GIM mechanism is already incorporated and
the up-quark contribution is understood to be subtracted from top- and
charm-loops. In particular then $X(x)$ has the property $X(0)=0$.
\hfill\break
It is a characteristic feature of the \kpnn decay amplitude that both top
and charm contributions are of comparable size, since the smallness of
$X(x_c)$ in comparison to $X(x_t)$ (due to $m_c\ll m_t$) is compensated
by the strong CKM suppression of the top contribution. By contrast this
suppression is absent e.g. for the CP-violating mode
$K_L\rightarrow \pi^o\nu\bar\nu$, where the branching ratio is determined
by the imaginary part of the amplitude, or for rare B-decays such as
$B\rightarrow X_s\nu\bar\nu$ and $B\rightarrow l^+l^-$. Therefore the
charm contribution is completely negligible in these latter processes.
\hfill\break
The short-distance electroweak loops of fig.1 receive calculable
radiative corrections through gluon exchange. Until recently these QCD
corrections have been known only to leading logarithmic (LLA) accuracy
\cite{EH,DDG,BBH}.
In this approximation the charm contribution to the \kpnn amplitude is
reduced by about 35\% through QCD effects. Since due to $m_t=O(M_W)$ no
large logs are present in the top contribution, this part remains
uncorrected in LLA. As a consequence a non-negligible theoretical
uncertainty remains due to the choice of the arbitrary renormalization
scale $\mu$ at which the top mass $m_t(\mu)$ is defined. Although the effect
of varying $\mu$ around values of $O(m_t)$ is formally of the neglected
order $O(\alpha_s)$, it amounts numerically to an uncertainty of 20\%--30\%
for branching ratios. This issue has been discussed in detail in our
previous papers \cite{BB1,BB2} where we have calculated
the complete $O(\alpha_s)$ corrections to the top
contribution for a class of rare decays governed by Z-penguin- and
box-diagrams. This allowed us to give improved expressions, including the
$O(\alpha_s)$ corrections, for the branching ratios of those decays that
are top-quark dominated ($K_L\rightarrow \pi^o\nu\bar\nu$,
$B\rightarrow X_s\nu\bar\nu$, $B\rightarrow l^+l^-$) and to show that the
scale ambiguities in the branching ratios are thus reduced from
$O(25\%)$ to typically $O(3\%)$.

\vspace{0.5cm}

The main objective of the present study is to perform a complete
renormalization group analysis of the charm contribution to \kpnn in the
next-to-leading logarithmic approximation (NLLA). Going beyond LLA is
mandatory for consistency with the QCD corrected top-quark contribution.
Furthermore it will decrease considerable theoretical uncertainties
due to the choice of $\mu=O(m_c)$ present also in the
charm sector. The remaining uncertainties will however be bigger than the
ones in the top-sector due to the smaller scales involved at which
$\alpha_s(\mu)$ is larger and
QCD perturbation theory less accurate. Combining the result with the QCD
corrected top contribution from \cite{BB2} will provide us with the final
theoretical expression for $B(\kpn)$ which includes the
short-distance QCD effects consistently at the next-to-leading-log level.

\vspace{0.5cm}

At this stage we would like to recall the strong dependence of
$B(\kpn)$ on the choice of $m_c$ as emphasized in particular by
Dib \cite{D} and by Harris and Rosner \cite{HR}. These authors varying
$m_c$ in the range $1.2 GeV\leq m_c\leq 1.8 GeV$ have found a 40-50\%
uncertainty in $B(\kpn)$ which is clearly disturbing. This large range
for $m_c$ chosen in \cite{D,HR} is at first sight surprising because in
a decay like \kpnn the presence of the charm quark is only felt in the
short distance loop and consequently it is evident that for $m_c$ the
current quark mass value should be used. For the latter however one has
$m_c(m_c)=1.3\pm 0.1 GeV$ \cite{GL,N}. On the other hand there is an
ambiguity in the choice of the scale in $m_c$ and the use of $m_c(\mu)$
with $\mu=O(m_c)$ in the calculation of $B(\kpn)$ is perfectly
legitimate. With $m_c(m_c)$ given above and $1 GeV\leq\mu\leq 3 GeV$
one effectively ends up with a range for $m_c(\mu)$ very similar to
the one chosen for $m_c$ in \cite{D,HR}. Consequently if one uses the
leading order expressions as done in \cite{D,HR}, the uncertainty due
to the treatment of $m_c$ is quite substantial. One of the important
results of the present analysis is that the inclusion of
next-to-leading corrections reduces considerably the uncertainty
due to $m_c$ stressed in \cite{D,HR}.

In order to further
motivate the present calculation it is instructive to
recall the relevant formulae for the top contribution \cite{BB2} and
discuss their implications for the charm case. The function $X(x)$ in
(\ref{hxxi}) is to $O(\alpha_s)$ and
without approximation in $m/M_W$ given by eqs.
 (\ref{xx})--(\ref{l2}).
In the case of charm ($x_c\approx 3\cdot 10^{-4}\ll 1$)
it is not necessary to keep
all orders in $m/M_W$. We will restrict ourselves to keeping only terms
linear in $x$, neglecting the order $O(x^2)$, which is an excellent
approximation. Indeed in this case one can write the function $X$ from
(\ref{xx}) with an error of less than 0.1\% as ($\mu=m_c$;
$a\equiv\alpha_s/4\pi$)
\b\l{xxc} X(x)\doteq -{3\o 4}x\ln x-{1\o 4}x+
     a\left(-2x\ln^2 x-7x\ln x-{23+2\pi^2\o 3}x\right)  \e
It is instructive to take a closer look at the numerical values of the
five terms in (\ref{xxc}). For typical values $m_c=1.3 GeV$,
$\Lambda_{QCD}=0.25 GeV$ these five terms read
\b\l{xxcn} X(x)=\left(16.32-0.66-13.05+5.54-1.37\right)\cdot 10^{-4}=
   6.78\cdot 10^{-4}  \e
Several lessons can already be learned from this simple exercise:
\begin{itemize}
\item
One can see very clearly that straightforward perturbation theory is not
reliable in this case as the $O(a)$ correction amounts to more than 50\%
of the lowest order result.
\item
A renormalization group treatment is therefore mandatory.
The LLA consists in summing
the first and third terms in (\ref{xxc}), (\ref{xxcn}) and similar terms
$O(x a^n \ln^{n+1} x)$ to all orders in perturbation theory.
\item
However the fourth term, neglected in LLA, is still quite sizeable. We can
see explicitly that a next-to-leading log summation is very desirable:
it resums all the terms of $O(x a^n \ln^n x)$, in particular the second
and the fourth term above.
\item
The non-leading mass term in the lowest order expression ($-{1\o 4}x$) is
consistently taken into account within NLLA.
\item
The suppression of the charm contribution by leading-log QCD corrections
is reflected in the negative sign of the $O(a x \ln^2 x)$ correction term.
The positive next-to-leading log contribution indicates that this
suppression tends to be weakened in NLLA. This will be confirmed by the
exact result discussed in section 5.
\item
The last term above is formally of the order $O(a x)$ and therefore
neglected even in NLLA. Unfortunately it has a relatively large coefficient
and gives roughly a 10\% contribution. This number gives a hint on the
remaining uncertainty to be expected.
\end{itemize}
\medskip
It is evident from this discussion that, given the high interest
in the rare decay mode \kpnn and
considering the theoretical situation, it is worthwhile to perform the
full next-to-leading order analysis in order to be able to reduce
disturbing theoretical uncertainties.\hfill\break
Our paper is organized as follows. Section 2 briefly collects
some of the most important results.
In sections 3 and 4 we will
describe in some detail the formalism necessary for the two-loop
renormalization group calculation in the charm sector for the Z-penguin-
and the box-part respectively. In section 5 the results will be combined
and the implications regarding the size of the corrections and the
dependence on the renormalization scale will be investigated. In
section 6 we combine both charm and top contributions and discuss
phenomenological consequences for \kpnn . In particular we address the
determination of $V_{td}$ from this decay and we discuss the
implications of our calculations for the unitarity triangle.
The calculation of next-to-leading order corrections to the decay
$K_L\rightarrow \mu^+\mu^-$ is very similar to \kpnn
and will be presented in section 7 for completeness.
Unfortunately, due to insufficiently known long distance contributions
this decay is less useful for the determination of CKM parameters.
We end our paper with a brief summary of the main results.

\section{The Main Formulae of this Paper}
Before entering the detailed discussion we find it useful to collect
in this section the most important results of the present paper.
\hfill\break
First of all, the complete effective hamiltonian relevant for \kpnn
can be written as
\b\l{hkpn} {\cal H}_{eff}={G_F \o{\sqrt 2}}{\alpha\o 2\pi \sin^2\Theta_W}
 \sum_{l=e,\mu,\tau}\left( V^{\ast}_{cs}V_{cd} X^l_{NL}+
V^{\ast}_{ts}V_{td} X(x_t)\right)
 (\bar sd)_{V-A}(\bar\nu_l\nu_l)_{V-A}  \e
The \kpnn branching ratio obtained from (\ref{hkpn}) is, for one single
neutrino flavor, given by
\b\l{bkpnl}
B(K^+\to\pi^+\nu_l\bar\nu_l)={\alpha^2 B(K^+\to\pi^oe^+\nu)\o
 V^2_{us} 2\pi^2 \sin^4\Theta_W}\left|V^\ast_{cs}V_{cd} X^l_{NL}+
 V^\ast_{ts}V_{td} X(x_t)\right|^2  \e
Here index $l$ ($l$=$e$, $\mu$, $\tau$) denotes the lepton flavor.
The lepton mass dependence in the top-quark sector can be
neglected. The function $X(x)$, relevant for the top contribution,
is to $O(\alpha_s)$ and to all orders in $m/M_W$ given by
($a\equiv \alpha_s/4\pi$)
\b\l{xx} X(x)=X_0(x)+a X_1(x) \e
with \cite{IL}
\b\l{xx0} X_0(x)={x\o 8}\left[ -{2+x\o 1-x}+{3x-6\o (1-x)^2}\ln x\right] \e
and the QCD correction \cite{BB1,BB2}
\begin{eqnarray}\l{xx1}
X_1(x)=&-&{23x+5x^2-4x^3\o 3(1-x)^2}+{x-11x^2+x^3+x^4\o (1-x)^3}\ln x
\nonumber\\
&+&{8x+4x^2+x^3-x^4\o 2(1-x)^3}\ln^2 x-{4x-x^3\o (1-x)^2}L_2(1-x)
\nonumber\\
&+&8x{\partial X_0(x)\o\partial x}\ln x_\mu
\end{eqnarray}
where $x_\mu=\mu^2/M^2_W$ with $\mu=O(m_t)$ and
\b\l{l2} L_2(1-x)=\int^x_1 dt {\ln t\o 1-t}   \e
The $\mu$-dependence in the last term in (\ref{xx1}) cancels to the
order considered the $\mu$-dependence of the leading term $X_0(x(\mu))$.
The corresponding expression for the charm sector is the function
$X^l_{NL}$. It results from the RG calculation in NLLA and reads as
follows:
\b\l{xlnl}X^l_{NL}=C_{NL}-4 B^{(1/2)}_{NL}  \e
\begin{eqnarray}\l{cnln}
\lefteqn{C_{NL}={x(m)\o 32}K^{{24\o 25}}_c\left[\left({48\o 7}K_++
 {24\o 11}K_--{696\o 77}K_{33}\right)\left({1\o a(\mu)}+{15212\o 1875}
 (1-K^{-1}_c)\right)\right.}\nonumber\\
&&+\left(1-\ln{\mu^2\o m^2}\right)(16K_+-8K_-)-{1176244\o 13125}K_+-
 {2302\o 6875}K_-+{3529184\o 48125}K_{33} \nonumber\\
&&+\left. K\left({56248\o 4375}K_+-{81448\o 6875}K_-+{4563698\o 144375}K_{33}
  \right)\right]
\end{eqnarray}
where
\b\l{kkc} K={a(M_W)\o a(\mu)}\qquad
  K_c={a(\mu)\o a(m)}  \e
\b\l{kkn} K_+=K^{{6\o 25}}\qquad K_-=K^{{-12\o 25}}\qquad
          K_{33}=K^{{-1\o 25}}  \e
\begin{eqnarray}\l{bnln}
\lefteqn{B^{(1/2)}_{NL}={x(m)\o 4}K^{24\o 25}_c\left[ 3(1-K_2)\left(
 {1\o a(\mu)}+{15212\o 1875}(1-K^{-1}_c)\right)\right.}\nonumber\\
&&-\left.\ln{\mu^2\o m^2}-
  {r\ln r\o 1-r}-{305\o 12}+{15212\o 625}K_2+{15581\o 7500}K K_2
  \right]
\end{eqnarray}
Here $K_2=K^{-1/25}$, $m=m_c$, $r=m^2_l/m^2_c$
and $m_l$ is the lepton mass.
We will at times omit the index $l$ of $X^l_{NL}$.
In (\ref{cnln}) -- (\ref{bnln}) the scale $\mu=O(m_c)$.
The two-loop expression for $a(\mu)$ is given in (\ref{als}).
Again -- to the considered order -- the explicit $\ln(\mu^2/m^2)$
terms in (\ref{cnln}) and (\ref{bnln}) cancel the $\mu$-dependence of
the leading terms.
\hfill\break
Expressing CKM elements through the Wolfenstein parameters
$\lambda$, $A$, $\varrho$ and $\eta$ with $\lambda=0.22$ and summing
over the three neutrino flavors we finally find
\b\l{bkpnum}
B(K^+\to\pi^+\nu\bar\nu)= 4.62\cdot 10^{-11}
  A^4 X^2(x_t)
 \left[ \eta^2+{2\o 3}(\varrho^e_o-\varrho)^2+{1\o 3}
 (\varrho^\tau_o-\varrho)^2 \right] \e
 where
\b\l{rh0lnum}
\varrho^l_o=1+{417\o A^2}{X^l_{NL}\o X(x_t)} \e
Formula (\ref{bkpnum}) together with $X(x_t)$ and $X^l_{NL}$ given in
(\ref{xx}) and (\ref{xlnl}) respectively gives $B(\kpn)$ in the standard
model with QCD effects taken into account consistently to next-to-leading
order. The leading order expressions are obtained by replacing
$X(x_t)\to X_0(x_t)$ and $X^l_{NL}\to X_L$ with $X_L$ found from
(\ref{cnln}) and (\ref{bnln}) by retaining there only the $1/a(\mu)$
terms. In LLA the one-loop expression should be used for $\alpha_s$.
This amounts to setting $\beta_1=0$ in (\ref{als}).
A numerical analysis of (\ref{bkpnum}) and (\ref{rh0lnum})
together with related issues will be presented in section 6.
Numerical values of $X^l_{NL}$ can be found in section 5.

\section{RG Calculation for the Z-Penguin Contribution}
In the present section we will outline the basic formalism needed to perform
the renormalization group analysis to next-to-leading order in the charm
sector of the decay $\kpn$. We start our discussion with the Z-penguin
contribution which is slightly more complicated than the box contribution.
Having derived the former, which can be done in closed form, it will be
staightforward to obtain the expression for the box contribution in the
next section.\hfill\break
The calculation consists of the following three steps:
\begin{itemize}
\item
In the first step the W and Z bosons being much heavier than the charm-quark
are integrated out and the resulting effective hamiltonian is written
down with matching conditions set up at $\mu=M_W$.
\item
The renormalization and mixing of the operators contributing to the
$\bar sd\rightarrow\bar\nu\nu$ transition amplitude is calculated to
two-loop order. The corresponding renormalization group equation is
solved and the scale $\mu$ is evolved down to lower values.
\item
At a scale $\mu=O(m_c)$ the charm quark as an explicit degree of freedom is
removed from the effective theory. The only remaining operator is
$(\bar sd)_{V-A}(\bar\nu\nu)_{V-A}$. Its coefficient yields the
generalization in NLLA of the charm penguin contribution to $X(x_c)$.
Since this operator has no anomalous dimension, no renormalization group
running occurs below scales of $O(m_c)$ and the contributing QCD
corrections are entirely perturbative.
\end{itemize}
For the actual calculations we took external quarks massless and
on-shell, which is the most convenient possibility.
Since we are computing the
short distance coefficients of an operator product expansion, the
treatment of external lines is arbitrary and the final result does not
depend on this choice. Furthermore we used dimensional
regularization for both UV- and IR-divergences. We have employed
Feynman gauge for the gluon and the W-boson field throughout the
present work.

\subsection{Step 1}
After the W and Z bosons have been integrated out, the effective
hamiltonian responsible for the Z-penguin contribution may be written as
($\lambda_c=V^{\ast}_{cs}V_{cd}$)
\b\l{hzop}{\cal H}^{(Z)}_{eff,c}={G_F \o{\sqrt 2}}
{\alpha\o 2\pi \sin^2\Theta_W}\lambda_c{\pi^2\o 2 M^2_W}
\left( v_+ O_+ +v_- O_- +v_3 Q\right) \e
where the operator basis is
\b\l{o1} O_1=
   -i\int d^4x\ T\left((\bar s_ic_j)_{V-A}(\bar c_jd_i)_{V-A}\right)(x)\
       \left((\bar c_kc_k)_{V-A}(\bar\nu\nu)_{V-A}\right)(0)\ -
       \{c\rightarrow u\}    \e
\b\l{o2} O_2=
   -i\int d^4x\ T\left((\bar s_ic_i)_{V-A}(\bar c_jd_j)_{V-A}\right)(x)\
       \left((\bar c_kc_k)_{V-A}(\bar\nu\nu)_{V-A}\right)(0)\ -
       \{c\rightarrow u\}    \e
\b\l{opm} O_\pm ={1\o 2}(O_2 \pm O_1) \e
\b\l{qnu} Q={m^2\o g^2} (\bar sd)_{V-A}(\bar\nu\nu)_{V-A}   \e
For convenience we have introduced the bilocal structures $O_{1,2}$, which
contain the relevant local 4-fermion operators in a form suitable for the
evaluation of the $\bar sd\rightarrow\bar\nu\nu$ transition amplitude.
Furthermore we have switched to the diagonal basis $O_\pm$, anticipating
the QCD renormalization and we have included a factor $m^2/g^2$
(where $m$ is the (charm-) quark mass and $g$ the strong coupling constant)
in the definition of $Q$ in order to streamline the notation for the
anomalous dimension matrices to be discussed below.
\hfill\break
As the first step we have to find the Wilson coefficients at $M_W$ to
next-to-leading order. These can be written as follows
($\vec v^T\equiv(v_+,v_-,v_3)$)
\b\l{vmw} \vec v(M_W)=\vec v^{(0)}+a(M_W)\vec v^{(1)}  \e
\b\l{v0} {\vec v^{(0)T}}=(1,1,0)  \e
\b\l{v1} {\vec v^{(1)T}}=(B_+,B_-,B_3)  \e
In the NDR-scheme ($\overline{MS}$, anticommuting $\gamma_5$
in $D\not=4$ dimensions) one gets
\b\l{bpm3} B_\pm=\pm 11{N\mp 1\o 2N}\qquad B_3=16  \e
where $N$ is the number of colors.
The calculation of the scheme dependent next-to-leading order terms
$B_\pm$
has been discussed in detail in \cite{BW}. The value of $B_3$ as well as
the zero entry in (\ref{v0}) follows from matching the
$\bar sd\rightarrow\bar\nu\nu$ amplitude calculated in the effective theory
(\ref{hzop}) onto the amplitude in the full theory to $O(a)$ at $\mu=M_W$.
To this end the renormalized matrix elements of the operators $O_\pm$
are needed. They correspond to the diagram of fig.2 and read in the
NDR-scheme
\b\l{mopm} \langle O_\pm\rangle =a(\mu)\ {1\o 2}\gamma^{(0)}_{\pm 3}
     \left(1-\ln{\mu^2\o m^2}\right) \langle Q\rangle  \e
with $\gamma^{(0)}_{\pm 3}$ given in (\ref{gzij}).
Using (\ref{mopm}) and comparing the matrix element of (\ref{hzop}) with
the amplitude in the full theory
\b\l{azc} A^{(Z)}_c ={G_F \o{\sqrt 2}}{\alpha\o 2\pi \sin^2\Theta_W}
 \lambda_c \left({x\o 4}\ln x+{3\o 4}x\right)
   (\bar sd)_{V-A}(\bar\nu\nu)_{V-A}  \e
gives the value of $B_3$ quoted in (\ref{bpm3}). The amplitude
$A^{(Z)}_c$ follows simply from the evaluation of the $Z^o$-penguin
diagrams of fig.1 in the limit $x\ll 1$.

\subsection{Step 2}
In the next step we have to discuss the renormalization group evolution.
The anomalous dimensions for $O_\pm$, to be denoted by $\gamma_\pm$,
are known to two-loop order and have
been calculated in \cite{BW,ACMP} in various renormalization schemes.
Here we use the NDR result of \cite{BW}.
The anomalous dimension of $Q$,
denoted by $\gamma_{33}$, is also
known to the same order. It depends only on the anomalous dimensions of
the mass and the coupling constant as is evident from the definition
(\ref{qnu}). What remains is the mixing of $O_\pm$ to $Q$
($\gamma_{\pm 3}$). The corresponding
elements of the anomalous dimension matrix are obtained as usual from the
divergent parts of the diagrams in figs. 2 and 3. The one-loop mixing has
been considered in \cite{EH,DDG,BBH}. The two-loop case required for the
present investigation is calculated in the present work and the results
will be given below. In short, the anomalous dimension matrix has in the
operator basis $\{ O_+,O_-,Q\}$ the general form
\b\l{gz} \gamma =
 \left(\begin{array}{ccc} \gamma_+ & 0 & \gamma_{+3} \\
                           0 & \gamma_- & \gamma_{-3} \\
                           0 & 0 & \gamma_{33}
    \end{array}\right)   \e
It governs the evolution of the Wilson coefficient functions by means of
the renormalization group equation
\b\l{rgv} {d\o d\ln\mu}\vec v=\gamma^T\vec v  \e
Note in particular that the evolution of $v_\pm$ is unaffected by the
presence of Q.\hfill\break
To proceed we briefly recall the general method for the solution of the
renormalization group equation (\ref{rgv}) in next-to-leading order, which
we find useful also for the special case at hand. We follow \cite{BJLW}.
\hfill\break
The anomalous dimensions of the operators $\{ O_\pm,Q\}$,
$\gamma$, and of
the coupling constant, $\beta(g)/g$, are in second order of perturbation
theory
\b\l{ga2} \gamma(a)=a \gamma^{(0)}+a^2\gamma^{(1)}  \e
\b\l{bg2} \beta(g)/g=-a\beta_0 -a^2\beta_1   \e
We also need the anomalous dimension of the mass
\b\l{rgm} {dm\o d\ln\mu}=-\gamma_m m  \e
\b\l{gma2} \gamma_m(a)=a \gamma_{m0}+a^2\gamma_{m1}  \e
The solution of (\ref{rgv}) can be expressed in terms of an evolution
matrix $U(\mu,M_W)$ as
\b\l{srg} \vec v (\mu)=U(\mu,M_W)\vec v(M_W)  \e
\b\l{umw} U(\mu,M_W)=[1+a(\mu) J]U^{(0)}(\mu,M_W)[1-a(M_W) J]  \e
$U^{(0)}$ is the evolution matrix in leading logarithmic approximation:
\b\l{umw0} U^{(0)}(\mu,M_W)=
  V\left({\left[{a(M_W)\o a(\mu)}\right]}^{{\vec\gamma^{(0)}\o 2\beta_0}}
   \right)_D V^{-1}   \e
where $V$ diagonalizes ${\gamma^{(0)T}}$
\b\l{g0d} \gamma^{(0)}_D=V^{-1} {\gamma^{(0)T}} V  \e
and $\vec\gamma^{(0)}$ is the vector containing the diagonal elements of
the diagonal matrix $\gamma^{(0)}_D$.\hfill\break
Defining now
\b\l{gdef} G=V^{-1} {\gamma^{(1)T}} V   \e
and a matrix $S$ whose elements are
\b\l{sij} S_{ij}=\delta_{ij}\gamma^{(0)}_i{\beta_1\o 2\beta^2_0}-
    {G_{ij}\o 2\beta_0+\gamma^{(0)}_i-\gamma^{(0)}_j}  \e
the matrix $J$ is given by
\b\l{jdef} J=V S V^{-1}   \e
In our case the $3\times 3$ matrix $J$ has the structure
\b\l{jmf} J =
 \left(\begin{array}{ccc}  J_{11} & 0 & 0 \\
                           0 & J_{22} & 0 \\
                           J_{31} & J_{32} & J_{33}
    \end{array}\right)   \e
similar to $\gamma^T$ in (\ref{gz}).\hfill\break
We will now collect various expressions which enter the renormalization
group functions discussed above.\hfill\break
For the QCD coupling in NLLA we use the form
\b\l{als}
a(\mu)={1\o{\beta_0\ln{{\mu^2}\o{\Lambda^2}}}} \left[1-
 {{\beta_1}\o{\beta_0^2}}{{\ln\ln{{\mu^2}\o{\Lambda^2}}}\o
    {\ln{{\mu^2}\o{\Lambda^2}}}}\right] \e
where $\Lambda\equiv\Lambda^{(4)}_{\overline{MS}}$. The parameters of the
$\beta$-function and of the anomalous dimension of the mass are
(with $N$($f$) the number of colors (flavors))
\b\l{bet}
\beta_0={{11N-2f}\o 3}\qquad
\beta_1={34\o 3}N^2-{10\o 3}Nf-2C_F f\qquad
C_F={{N^2-1}\o{2N}}\e
\b\l{gm01} \gamma_{m0}=6C_F\qquad \gamma_{m1}=C_F\left(
     3C_F+{97\o 3}N-{10\o 3}f\right)  \e
The nonvanishing elements of the anomalous dimension matrix read
\b\l{gzij}
\begin{tabular}{lcl}
$\gamma^{(0)}_{33}=2(\gamma_{m0}-\beta_0)$ &\qquad\qquad&
   $\gamma^{(1)}_{33}=2(\gamma_{m1}-\beta_1)$ \\  & & \\
$\gamma^{(0)}_\pm=\pm 6{N\mp 1\o N}$ &  &
   $\gamma^{(1)}_\pm={N\mp 1\o 2N}\left(-21\pm{57\o N}\mp{19\o 3}N\pm{4\o 3}f
     \right)$ \\                                 & & \\
$\gamma^{(0)}_{\pm 3}=\pm 8(N\pm 1)$ &  &
   $\gamma^{(1)}_{\pm 3}=C_F(\pm 88 N-48)$ \\
\end{tabular}  \e
The expressions $\gamma^{(1)}$ are in the NDR-scheme. The matrix
$V$ in (\ref{g0d}) is
\b\l{vvi} V =
 \left(\begin{array}{ccc}  1 & 0 & 0 \\
                           0 & 1 & 0 \\
                           V_{31} & V_{32} & 1
    \end{array}\right)   \e
\b\l{v312} V_{31}={\gamma^{(0)}_{+3}\o \gamma^{(0)}_+-\gamma^{(0)}_{33}}
\qquad V_{32}={\gamma^{(0)}_{-3}\o \gamma^{(0)}_--\gamma^{(0)}_{33}} \e
Furthermore we define (see (\ref{kkc}))
\b\l{kkk} K_i=K^{\gamma^{(0)}_i\o 2\beta_0}\qquad {i=+,-,33}  \e
With all these formulae at hand we are now able to obtain the coefficients
$\vec v$ at scales $\mu\ll M_W$ by means of (\ref{srg}). We find
\b\l{vpme}v_\pm(\mu)=[1+a(\mu) J_\pm]\ K_\pm\ [1-a(M_W) (J_\pm-B_\pm)]\e
where $J_+\equiv J_{11}$, $J_-\equiv J_{22}$ and
\begin{eqnarray}\l{v3e}
\lefteqn{v_3(\mu)=V_{31}(K_+-K_{33}) + V_{32}(K_--K_{33})}\nonumber\\
&&+a(\mu)\left[(J_{31}+J_{33}V_{31})K_+ +(J_{32}+J_{33}V_{32})K_- -
   J_{33}(V_{31}+V_{32}) K_{33}\right]\nonumber\\
&&+a(M_W)\left[(B_+-J_{11})V_{31}K_++ (B_--J_{22})V_{32}K_-\right.\nonumber\\
&&\qquad +\left.
  \left(B_3-J_{31}-J_{32}+(J_{11}-B_+)V_{31}+(J_{22}-B_-)V_{32}\right)K_{33}
  \right]
\end{eqnarray}

\subsection{Step 3}
The last step consists in integrating out the charm-quark. This is done
by matching the effective theory in (\ref{hzop}) onto an effective theory
without the operators $O_\pm$ at a scale $\mu=O(m_c)$. Below this scale
the hamiltonian involves the single operator $Q$. The effects of
the charm field being removed from the theory as an explicitly appearing
degree of freedom are then still indirectly present in the modified
coefficient of $Q$. Effectively this step is accomplished by inserting
$\langle O_\pm\rangle$ of (\ref{mopm}) into (\ref{hzop}).

\subsection{Final Result for the Z-Penguin}
Collecting the results of the three steps and using (\ref{hzop})
we obtain the final
expression in NLLA for the charm contribution to the Z-penguin part of
the effective hamiltonian, suitable for scales below $\mu=O(m_c)$:
\b\l{hzc} {\cal H}^{(Z)}_{eff,c}={G_F \o{\sqrt 2}}
  {\alpha\o 2\pi \sin^2\Theta_W} \lambda_c\ C_{NL}\
   (\bar sd)_{V-A}(\bar\nu\nu)_{V-A}  \e
\b\l{cnl} C_{NL}={x(\mu)\o 32}\left[{1\o 2}\left(1-\ln{\mu^2\o m^2}\right)
  \left(\gamma^{(0)}_{+3} K_++\gamma^{(0)}_{-3} K_-\right)+
  {v_3(\mu)\o a(\mu)}\right]  \e
To the order considered the terms of $O(a)$ in $v_\pm(\mu)$
have to be dropped.
$C_{NL}$ is the generalization of the lowest order function $C_0(x)$
(see (\ref{lc0})) that includes the QCD corrections in NLLA. In fact
\b\l{lc0} \lim_{\Lambda\to 0} C_{NL}={x\o 4}\ln x+{3\o 4}x\doteq C_0(x) \e
We will discuss this result in more detail later on.\hfill\break
It is useful to express the running charm quark mass $m(\mu)$ in terms of
$m(m)$, which we use as an input parameter:
\b\l{xmu} x(\mu)=x(m)\left[{a(\mu)\o a(m)}\right]^{{\gamma_{m0}\o \beta_0}}
 \left[1+\left({\gamma_{m1}\o\beta_0}-{\beta_1\gamma_{m0}\o\beta^2_0}
 \right)\left(a(\mu)-a(m)\right)\right]  \e
Setting $N=3$ and $f=4$ in (\ref{cnl})
we then obtain formula (\ref{cnln})
quoted in section 2.\hfill\break
We have performed the renormalization group evolution in an effective
four-flavor theory from scales of $O(M_W)$ down to $O(m_c)$ and
neglected the effects of a b-quark threshold and a corresponding
five-flavor theory above $m_b$. This is an excellent approximation
which allows one to write down a fairly compact final result in
closed form. We have checked that the inclusion of a b-quark
threshold would change the values for the charm-quark function
$X_{NL}$ by not more than 0.1\%.

\subsection{Renormalization Scheme Independence}
To conclude this section we turn to the question of renormalization scheme
dependences and their cancellation. This issue is well known and has
already been discussed elsewhere (see e.g. \cite{BJLW}). However since
the application of the next-to-leading renormalization group formalism to
the case considered here is new and the explicit expression of $C_{NL}$
clearly involves scheme dependences whose cancellation may not be entirely
obvious from (\ref{v3e}), we find it useful to give a short account of this
subject in explicit terms.\hfill\break
In essence the scheme independence of a matrix element of the effective
hamiltonian (\ref{hzop}) is trivial since it is proportional to
\b\l{ovmu} \langle\vec O^T(\mu)\rangle\vec v(\mu)=
 \langle v_+O_++v_-O_-+v_3 Q\rangle\vert_\mu \e
which is manifestly scheme independent. Different choices of the scheme
just shift finite $O(a)$-terms between the coefficients and the operator
matrix elements--this freedom is in fact the reason why the unphysical
coefficients are scheme dependent. More explicitly these quantities are in
different schemes related by
\b\l{ovp}\langle\vec O\rangle'=(1-a s)\langle\vec O\rangle\qquad
 \vec v'=(1+a s^T)\vec v  \e
 where $s$ is a constant $3\times 3$ matrix of the form
\b\l{smf} s =
 \left(\begin{array}{ccc}  s_+ & 0 & s_{+3} \\
                           0 & s_- & s_{-3} \\
                           0 & 0 & s_3
    \end{array}\right)   \e
Since
\b\l{oumv} \langle\vec O^T(\mu)\rangle\vec v(\mu)\equiv
  \langle\vec O^T(\mu)\rangle U(\mu,M_W) \vec v(M_W)  \e
implies
\b\l{uup} U'(\mu,M_W)=(1+a(\mu)s^T)U(\mu,M_W)(1-a(M_W)s^T)  \e
we have from (\ref{umw})
\b\l{jjp} J'=J+s^T \e
If we use the relation between $J$ and $\gamma^{(1)}$ (from (\ref{gdef}) -
(\ref{jdef}))
\b\l{jg1} J+\left[{\gamma^{(0)T}\o 2\beta_0},J\right]=
  -{\gamma^{(1)T}\o 2\beta_0}+{\beta_1\o 2\beta^2_0}\gamma^{(0)T} \e
we obtain the transformation property of the anomalous dimensions under
a change in the renormalization scheme:
\b\l{ggp} \gamma^{(1)\prime}=\gamma^{(1)}-[s,\gamma^{(0)}]-2\beta_0 s\qquad
   \gamma^{(0)\prime}=\gamma^{(0)}  \e
Given these general formulae one can easily clarify the cancellation of
scheme dependences in all particular cases or convert scheme dependent
quantities from one scheme to another, if desired. To be more specific,
let us consider expression (\ref{v3e}). The first line being a function
of one-loop anomalous dimensions and the coupling constant only is
obviously scheme independent. The coefficient of $a(M_W)$
is scheme independent
since the scheme dependences in $J$ and $B_i$ cancel in the
combinations $B_+-J_{11}$, $B_--J_{22}$ and $B_3-J_{31}-J_{32}$
((\ref{v1}), (\ref{ovp}) and (\ref{jjp})).
The scheme dependences in the coefficient of $a(\mu)$
remain and are responsible for the scheme dependence of $v_3(\mu)$.
We have, using (\ref{v3e}), (\ref{jjp}) and
$\langle Q\rangle'=(1-a s_3)\langle Q\rangle$
\b\l{v3qp}
v'_3\langle Q\rangle'=v_3\langle Q\rangle +a(\mu)
[s_{+3}K_++s_{-3}K_-]\langle Q\rangle  \e
On the other hand, up to terms of the neglected order (note that
$\langle O_{\pm}\rangle=O(a)$)
\b\l{vpmop}
v'_+\langle O_+\rangle'+v'_-\langle O_-\rangle'
=v_+\langle O_+\rangle +v_-\langle O_-\rangle
-a(\mu) [s_{+3}K_++s_{-3}K_-]\langle Q\rangle  \e
Eq. (\ref{vpmop}) reflects the arbitrariness one has in the subtraction
of finite constants when defining the renormalized matrix elements
$\langle O_{\pm}\rangle$. The scheme dependences in (\ref{v3qp}) and
(\ref{vpmop}) cancel in the sum, which essentially represents the
function $C_{NL}$.\hfill\break
The cancellation of the $\mu$-dependence will be discussed in section 5.

\section{RG Calculation for the Box Contribution}
After the detailed exposition of the Z-penguin contribution in the preceding
section it is straightforward to repeat the analysis in the analogous, but
somewhat simpler case of the box. We will briefly summarize the essential
steps.
\subsection{Step 1}
When the W boson is integrated out, the hamiltonian relevant for the box
diagram is given by
\b\l{hbop}{\cal H}^{(B)}_{eff,c}=-{G_F \o{\sqrt 2}}
{\alpha\o 2\pi \sin^2\Theta_W}\lambda_c\left(-{\pi^2\o M^2_W}\right)
\left( c_1 O +c_2 Q\right) \e
\b\l{ob} O=
   -i\int d^4x\ T\left((\bar sc)_{V-A}(\bar \nu l)_{V-A}\right)(x)\
       \left((\bar l\nu)_{V-A}(\bar cd)_{V-A}\right)(0)\ -
       \{c\rightarrow u\}    \e
\b\l{qnub} Q={m^2\o g^2} (\bar sd)_{V-A}(\bar\nu\nu)_{V-A}   \e
In order to find the initial conditions for the coefficients $c_i$
we need the renormalized matrix element of the operator $O$. Calculating
the diagram of fig.4 we find in the NDR-scheme
\b\l{mob} \langle O\rangle=a(\mu) 16\left(\ln{\mu^2\o m^2}+{5\o 4}+
  {r \ln r\o 1-r}\right)\langle Q\rangle  \e
Inserting this result into (\ref{hbop}) and
comparing with the amplitude in
the full theory given by
\b\l{abc} A^{(B)}_c =-{G_F \o{\sqrt 2}}{\alpha\o 2\pi \sin^2\Theta_W}
 \lambda_c \left(x\ln x+x\left(1+{r\ln r\o r-1}\right)\right)
   (\bar sd)_{V-A}(\bar\nu\nu)_{V-A}  \e
implies the initial
values for the Wilson coefficient function in the NDR-scheme
\b\l{cmw} \vec c^T(M_W)\equiv (c_1(M_W),c_2(M_W))=(1,0)+a(M_W)(0,B_2)
  \qquad B_2=-36  \e
\subsection{Step 2}
The anomalous dimension matrix in the basis $\{O,Q\}$ has the form
\b\l{gb} \gamma =
 \left(\begin{array}{cc}   0 & \gamma_{12} \\
                           0 & \gamma_{22}
    \end{array}\right)   \e
with the perturbative expansion
\b\l{gb2} \gamma=a \gamma^{(0)}+a^2\gamma^{(1)}  \e
The anomalous dimensions $\gamma^{(0)}_{12}$, $\gamma^{(1)}_{12}$, describing
the RG mixing of operator $O$ with $Q$ follow from the divergent parts of
the diagrams in fig.4 and fig.5 respectively.
$\gamma^{(0)}_{12}$ has been calculated in \cite{EH,DDG,BBH}.
$\gamma^{(1)}_{12}$ is new.
The solution of the
renormalization group equation in NLLA is obtained via the general
method described in section 3.\hfill\break
The matrix $V$ needed to diagonalize $\gamma^{(0)T}$
\b\l{vgvb} V^{-1}\gamma^{(0)T} V=
 \left(\begin{array}{cc}   0 & 0 \\
                           0 & \gamma^{(0)}_{22}
    \end{array}\right)   \e
is given by
\b\l{vvb} V=
 \left(\begin{array}{cc}   1 & 0 \\
                           V_{21}  & 1
    \end{array}\right) \qquad V_{21}=-{\gamma^{(0)}_{12}\o
                     \gamma^{(0)}_{22}}   \e
Furthermore, the matrix J defined in (\ref{jdef}) can in this case be
written as
\b\l{jbmf} J=
 \left(\begin{array}{cc}   0 & 0 \\
                           J_{21} & J_{22}
    \end{array}\right)   \e
The nonvanishing elements of the anomalous dimension matrix are
\b\l{gbij}
\begin{tabular}{lcl}
$\gamma^{(0)}_{22}=2(\gamma_{m0}-\beta_0)$ &\qquad\qquad&
   $\gamma^{(1)}_{22}=2(\gamma_{m1}-\beta_1)$ \\  & & \\
$\gamma^{(0)}_{12}=-32$ &  &
   $\gamma^{(1)}_{12}=80C_F$ \\
\end{tabular}  \e
Defining $K_2=K^{\gamma^{(0)}_{22}/2\beta_0}$ we then find for the
solution of the renormalization group equation
\b\l{rgc} {d\o d\ln\mu}\vec c=\gamma^T\vec c  \e
\b\l{c1e} c_1(\mu)\equiv 1 \e
\b\l{c2e} c_2(\mu)=V_{21}(1-K_2)+a(\mu)\left[J_{21}+V_{21}J_{22}
  (1-K_2)\right] +a(M_W) (B_2-J_{21})K_2  \e
\subsection{Step 3}
The last step consists of integrating out the charm quark. This adds
an additional contribution to the coefficient of the operator $Q$
which in view of (\ref{c1e}) is simply given by the renormalized
matrix element of the operator $O$ at a scale $\mu=O(m)$, already
quoted in (\ref{mob}).
\subsection{Final Result for the Box Contribution}
Collecting all this and using (\ref{hbop})
we finally obtain the effective hamiltonian induced from box diagrams
with charm, including the NLL QCD effects between the scales $M_W$ and
$\mu=O(m_c)$:
\b\l{hbcn} {\cal H}^{(B)}_{eff,c}=-{G_F \o{\sqrt 2}}
  {\alpha\o 2\pi \sin^2\Theta_W} \lambda_c\ 4 B^{(1/2)}_{NL}\
   (\bar sd)_{V-A}(\bar\nu_l\nu_l)_{V-A}  \e
\b\l{bnl} B^{(1/2)}_{NL}=-{x(\mu)\o 64}\left[ 16\left(\ln{\mu^2\o m^2}+
    {5\o 4}+{r\ln r\o 1-r}\right)+{c_2(\mu)\o a(\mu)}\right]  \e
(\ref{hbcn}) is written here for one $\nu$-flavor. The index $(1/2)$
refers to the weak isospin of the final state leptons.\hfill\break
If we set $N=3$, $f=4$ and use (\ref{xmu})
we arrive at the expression given in (\ref{bnln}).\hfill\break
We would like to add a few important remarks.
\begin{itemize}
\item
As usual we are working in the NDR-scheme; all scheme
dependent quantities that we have given above (one-loop initial values
and two-loop anomalous dimensions) refer to this scheme. However in the
case of the box contribution there is an additional scheme dependence
involved which arises from the treatment of the particular Dirac
structure. In our calculation we have used the following projection
\b\l{pjct} \gamma^\mu\gamma^\alpha\gamma^\nu(1-\gamma_5)\otimes
  \gamma_\mu\gamma_\alpha\gamma_\nu(1-\gamma_5)\to  4(4-
 \varepsilon) \gamma^\mu(1-\gamma_5)\otimes\gamma_\mu(1-\gamma_5)
 \e
A different treatment of this expression would lead to different
values for $B_2$ and $\gamma^{(1)}_{12}$ and would correspond to a
different renormalization scheme. This "projection scheme dependence"
of course cancels in the final result provided coefficient and
anomalous dimension are evaluated consistently.
\item
The proof of the cancellation of all scheme dependences (e.g. the
projection scheme dependence mentioned above) proceeds in the same
way as discussed at the end of section 3.
\item
The operator $O$ contains quark fields only in the form of local
currents and therefore has no anomalous dimension. This feature is
reflected in the structure of the anomalous dimension matrix in
(\ref{gb}). As a consequence the coefficient $c_1$ of $O$ is
$\mu$-independent (\ref{c1e}).
\item
In contrast to the case of the Z-penguin the box contribution involves
an additional mass parameter $m_l$, the mass of the internal charged
lepton in the box diagram. While $m_l$ can be neglected compared to the
charm mass ($r\approx 0$) in the case of the electron or the muon, it
is certainly non-negligible in the case of the tau. The effect of a
nonvanishing lepton mass of the order of the charm quark mass is
represented by the $r$-dependent term in the expressions for
$B^{(1/2)}_{NL}$
in (\ref{bnl}), (\ref{bnln}). A comparison with (\ref{mob}) shows that
the dependence on $m_l$ arises entirely from the matrix element of the
operator $O$, evaluated at the scale $\mu=O(m_c)$. Note in
particular that the initial values for the Wilson coefficients
(\ref{cmw}) and the anomalous dimensions (\ref{gbij}) do not depend on
$m_l$. This is intuitively clear since the Wilson coefficient depends
only on the short distance structure of the effective interaction and
must therefore be independent of the low energy scale $m_l$.\hfill\break
In this context the GIM mechanism plays an essential role. The
contracted box diagrams with charm (figs.4, 5) obviously have a quadratic
divergence. However, what really matters for the matrix element of $O$
is the {\it difference} between the charm- and the up-quark contribution
(see (\ref{ob})). As a consequence the matrix element
$\langle O\rangle$ is only logarithmically divergent. It follows then
that the difference between the cases $m_l=0$ and $m_l\not= 0$ is
{\it finite} and hence does not affect the anomalous dimensions, as
it must be in a mass independent scheme like minimal subtraction.
To conclude, the effect of the GIM mechanism is nontrivial in the case
of the box with nonvanishing lepton mass, whereas it is not explicitly
visible for $m_l=0$ (and in the case of the Z-penguin) since in
these latter cases the up-quark contribution is anyway vanishing in
dimensional regularization.
\end{itemize}

\section{Discussion of the Charm Con\-tri\-bu\-tion \hfill\break to
Next-\-To-\-Leading  Order}
In sections 3 and 4 we have calculated the Z-penguin and box-parts of
the charm contribution to the effective hamiltonian for \kpnn decay
in NLLA.
The explicit results for the functions $C_{NL}$ and $B^{(1/2)}_{NL}$
are given in (\ref{cnln}) and (\ref{bnln}) respectively.
Combining these functions, which depend on
the W-boson gauge, into the gauge independent combination
\b\l{xnl} X_{NL}=C_{NL}-4B^{(1/2)}_{NL}  \e
we can next write down the final result for the charm contribution
to the effective hamiltonian
\b\l{hcnl} {\cal H}_{eff,c}={G_F \o{\sqrt 2}}
  {\alpha\o 2\pi \sin^2\Theta_W} \lambda_c\ X_{NL}\
   (\bar sd)_{V-A}(\bar\nu\nu)_{V-A}  \e
We proceed to discuss various aspects of this result in what
follows.\hfill\break
Let us first compare the charm- with the top-contribution as obtained
in \cite{BB2}. In the case of the top the internal quark mass is large
and comparable to $M_W$ so that no large logarithms appear. The charm
mass on the other hand is very much smaller than $M_W$ and logarithmic
terms become important. For these reasons the two cases, regarding the
theoretical methods required, are quite complementary to each other:
For the top-part a consideration of QCD effects in straightforward
perturbation theory to $O(\alpha_s)$ is appropriate, but all orders have
to be kept in the mass ratio $m_t/M_W$. On the contrary the charm
contribution allows a restriction to the order $O(x_c)$ in the mass
ratio but requires summation of leading and next-to-leading logarithms
to all orders in QCD. From this consideration it follows that one can
check the top contribution against the charm contribution by taking
the limit $x\ll 1$ in the former case and simultaneously expanding the
RG result to $O(a)$ in the latter.\hfill\break
Recall that the function $X$ in (\ref{xx}) consists of a Z-penguin-
and a box-part $C$ and $B$, respectively
\b\l{xcb} X(x)=C(x)-4B(x,+1/2)  \e
Keeping only terms linear in $x$ in the exact results of \cite{BB2} we
have ($m_l=0$)
\b\l{cxc} C(x)\doteq{1\o 4}x\ln x+{3\o 4}x+a\left(2x\ln^2x+{23\o 3}
x\ln x+{29+2\pi^2\o 3}x\right) \e
\b\l{bnxc} 4B(x,+1/2)\doteq x\ln x+x
+a\left(4x\ln^2x+{44\o 3}
x\ln x+{52+4\pi^2\o 3}x\right) \e
Expanding now the RG-functions $C_{NL}$ and $B^{(1/2)}_{NL}$
(for $r=0$) in (\ref{cnln}) and (\ref{bnln}) to
$O(a)$, thereby "switching off" the renormalization group summations,
we obtain in fact the first four terms in (\ref{cxc}) and
(\ref{bnxc}) respectively. Let us mention a few interesting points.
\begin{itemize}
\item
The last terms in (\ref{cxc}), (\ref{bnxc}) are of $O(a x)$
and go beyond the NLLA; therefore they can of course not be obtained
from $C_{NL}$ and $B_{NL}$.
\item
A comparison of $B_{NL}$, expanded to $O(a)$, with (\ref{bnxc}) may
also be used to determine the two-loop anomalous dimension
$\gamma^{(1)}_{12}$. Once $\gamma^{(1)}_{12}$ is known, the full
next-to-leading order expression for the box can be obtained. This
observation allows to circumvent the explicit evaluation of the
two-loop diagrams in the effective theory (fig.5) -- the information
on $\gamma^{(1)}_{12}$ is already contained in the calculation of
$B(x,+1/2)$ in \cite{BB2}.
\item
By contrast, the same exercise for $C_{NL}$ shows that (\ref{cxc}) only
yields the value for ($\gamma^{(1)}_{+3}+\gamma^{(1)}_{-3}$). To get
both elements $\gamma^{(1)}_{\pm 3}$ separately it is necessary to
investigate the calculation in the effective theory (fig.3). Still
this feature can be used as a cross-check.
\end{itemize}
An important issue is the independence of the physical amplitude on the
renormalization scale $\mu$. Since in our case the operator
$(\bar sd)_{V-A}(\bar\nu\nu)_{V-A}$ has no anomalous dimension, also
the coefficient functions $C$, $B$ and $X$ are $\mu$-independent.
However, because we are working in perturbation theory, the
$\mu$-independence is valid only up to terms of the neglected order.
As a side remark we mention that the renormalization scale $\mu$
appearing in our final expressions $C_{NL}$ and $B_{NL}$ can be viewed
also as the scale at which the charm quark is removed from the
effective theory. This interpretation is clear from step 3 of the
derivation given in section 3. Although this scale should be of the same
order as the charm quark mass, it is not necessary that $\mu$ equals
$m_c$ exactly. Therefore we have displayed the $\mu$-dependence
explicitly in our formulae and have not set $\mu=m_c$. This allows us
in particular to investigate the residual $\mu$-dependence. To do so
consider first the leading-log parts of $C_{NL}$ and $B_{NL}$
(in (\ref{cnl}) and (\ref{bnl})) and their variation with $\mu$:
\begin{eqnarray}\l{mdpc}
\lefteqn{{d\o d\ln\mu}\left({x(\mu)\o 32 a(\mu)}\left[
V_{31}(K_+-K_{33}) + V_{32}(K_--K_{33})\right]\right)=}\hspace{4cm}
\nonumber\\
& &{x\o 32}\left[ \gamma^{(0)}_{+3}K_++\gamma^{(0)}_{-3}K_-\right]+
 O(ax)
\end{eqnarray}
\b\l{mdpb}
{d\o d\ln\mu}\left({x(\mu)\o 64 a(\mu)}
V_{21}(K_2-1)\right)=
-{x\o 64} \gamma^{(0)}_{12}+ O(ax)  \e
We can see that the residual $\mu$-dependence for the leading-log
result is of the order $O(x)$. Formally terms of this order are
neglected in LLA and the results are $\mu$-independent to the order
considered in this approximation. Numerically however the residual
$\mu$-dependence is quite pronounced as we will see below. Furthermore
we note again that treating QCD corrections in LLA and keeping the
non-logarithmic $O(x)$ terms in $X_0(x)$ would strictly speaking be
inconsistent; a variation of the scale $\mu$, which is not fixed in LLA,
would give rise to terms of the same order. Similarly the non-negligible
contribution from the tau-lepton mass term is a next-to-leading order
effect (see (\ref{bnl})).\hfill\break
All these considerations underline once more that the leading-log
approximation is unsatisfactory. An improvement can only be achieved
by going beyond the leading order. In particular in NLLA the
$\mu$-dependence of $O(x)$ in the leading order terms of $C_{NL}$ and
$B_{NL}$ given in (\ref{mdpc}) and (\ref{mdpb}) is cancelled by the
explicitly $\mu$-dependent terms appearing at next-to-leading order,
as is evident from (\ref{cnl}) and (\ref{bnl}). As a result the
remaining $\mu$-dependence is $O(ax)$ instead of $O(x)$. What this
amounts to numerically is illustrated in fig.6 where we have plotted
the charm-function $X_{NL}$ (for $m_l=0$)
compared to the leading-log result $X_L$
and the case without QCD for values of $\mu$ between
1 and 3 $GeV$. We observe the following features:
\begin{itemize}
\item
The residual slope of $X_{NL}$ is indeed considerably smaller in
comparison to $X_L$, which exhibits a quite substantial dependence
on the unphysical scale $\mu$.
The variation of $X$ (defined as $(X(1GeV)-X(3GeV))/X(m_c)$)
is 18.8\% in NLLA compared to 52.7\% in LLA.
\item
As already anticipated in the Introduction, the suppression of the
uncorrected function through QCD effects is somewhat less pronounced
in NLLA.
\item
The next-to-leading effects amount to a $\sim 10\%$ correction relative
to $X_L$ at $\mu=m_c$. However the size of this correction strongly
depends on $\mu$ due to the scale ambiguity of the result. This means
that the question of how large the next-to-leading effects compared to
the LLA really are cannot be answered uniquely. Therefore we focus our
attention on the more important issue of the $\mu$-dependence and its
reduction in NLLA.
\end{itemize}
Whereas it is meaningful to use $\Lambda_{\overline{MS}}$ in $X_{NL}$
one has no control on the exact definition of the QCD scale in
leading order ($\Lambda_{LO}$). $\Lambda_{LO}$ must be of the same
order of magnitude as $\Lambda_{\overline{MS}}$ but is undetermined
otherwise, which constitutes a further problem of the LLA. For
definiteness of the comparison we have set $\Lambda_{LO}=
\Lambda_{\overline{MS}}=0.25GeV$ in fig.6.
Finally in table 1 we give the numerical values for $X^e_{NL}$
and $X^{\tau}_{NL}$ for $\mu=m_c$ and several values of
$\Lambda_{\overline{MS}}$ and $m_c(m_c)$.
The reduction of $X_{NL}$ through the effect of $m_\tau$ is
clearly visible.
\begin{table}
\begin{center}
\begin{tabular}{|c|c|c|c|c|c|c|}\hline
&\multicolumn{3}{c|}{$X^e_{NL}/10^{-4}$}
&\multicolumn{3}{c|}{$X^{\tau}_{NL}/10^{-4}$}
  \\ \hline
$\Lambda_{\overline{MS}}$, $m_c$ [$GeV$]
&1.2&1.3&1.4&1.2&1.3&1.4\\ \hline
0.15&10.24&11.95&13.78&6.99&8.40&9.92\\ \hline
0.25&9.51&11.18&12.97&6.26&7.63&9.11\\ \hline
0.35&8.74&10.39&12.15&5.49&6.83&8.29\\ \hline
\end{tabular}
\end{center}
\centerline{}
{\bf Table 1:} The function $X_{NL}$ for various
$\Lambda_{\overline{MS}}$
and $m_c$.
\end{table}

\section{Phenomenological Implications for \hfill\break
         $\kpn$, $V_{td}$ and the Unitarity Triangle}

\subsection{General Expressions}
In the present section we will finally discuss the expression for
the branching ratio of \kpnn, including the complete $O(\alpha_s)$
corrections to the top contribution and the renormalization group
result in NLLA for the charm part. As applications we will discuss
consequences for a possible determination of $|V_{td}|$ from a
measured $B(\kpn)$ and the impact of our results on the unitarity
triangle. In our presentation we follow chapter 6.2. of
\cite{BH}. There a detailed investigation of the \kpnn phenomenology
has been performed with QCD corrections treated in LLA. While the
discussion in \cite{BH} concentrated on the various uncertainties
arising from standard model parameters like CKM elements and quark
masses, we will put emphasis on the theoretical uncertainties due to
residual dependences on the renormalization scale.\hfill\break
Taking (\ref{bkpnl}), summing over the three neutrino flavors and
expressing CKM variables in terms of the Wolfenstein parameters
$\lambda$, $A$, $\varrho$ and $\eta$, we obtain for the branching ratio
\begin{eqnarray}\l{bkpn}
\lefteqn{B(K^+\to\pi^+\nu\bar\nu)=} & &\nonumber\\
& & {3\alpha^2 B(K^+\to\pi^oe^+\nu)\o
 2\pi^2 \sin^4\Theta_W} \lambda^8 A^4 X^2(x_t)
 \left[ \eta^2+{2\o 3}(\varrho^e_o-\varrho)^2+{1\o 3}
 (\varrho^\tau_o-\varrho)^2 \right]
\end{eqnarray}
 where
\b\l{rh0l}
\varrho^l_o=1+{1-{\lambda^2\o 2}\o \lambda^4 A^2}{X^l_{NL}\o X(x_t)}\e
Using the numerical values for various parameters as collected in
(\ref{mwtl}) and (\ref{aswb}) we find the expressions quoted
in section 2.
Here and in the following we have used the QCD corrected top
contribution $X(x_t)$ as it stands and made no effort to expand
functions of $X(x_t)$ to first order in $\alpha_s$. For renormalization
scales not too far from $m_t$ this is anyway irrelevant
numerically.\hfill\break
Eq.(\ref{bkpn}) defines a circle in the $(\varrho,\eta)$ plane with
center at $(\varrho_o,0)$, where
\b\l{rhz}
\varrho_o=1+(1-{\lambda^2\o 2}){X^e_{NL}-{1\o 3}\delta X^\tau_{NL}\o
 \lambda^4 A^2 X(x_t)}   \e
and with the radius squared
\b\l{r02}
r^2_o={1\o \lambda^8 A^4 X^2(x_t)}\left[{2\pi^2\sin^4\Theta_W
B(\kpn)\o 3\alpha^2 B(K^+\to\pi^oe^+\nu)}-{2\o 9}
\left( (1-{\lambda^2\o 2})\delta X^\tau_{NL}\right)^2\right]  \e
Here we have defined
\b\l{delx}
\delta X^\tau_{NL}=X^e_{NL}-X^\tau_{NL}={x r\ln r\o r-1} \e
The restriction to leading logarithmic accuracy in the above formulae
is made by replacing $X(x_t)\to X_0(x_t)$ and $X^l_{NL}\to X_L$.
The difference $\delta X^\tau_{NL}$ should be set to zero in LLA for
consistency as discussed in section 5.\hfill\break
We recall a few important definitions related to the unitarity triangle
in the $(\varrho,\eta)$ plane. First we introduce \cite{BH}
\b\l{rbtd}
R_b={|V_{ub}|\o |V_{us}V_{cb}|}\qquad\qquad
R_t={|V_{td}|\o |V_{us}V_{ts}|} \e
To a very good approximation these can be expressed in terms of
$\varrho$ and $\eta$ as
\b\l{rbt}
R_b=\sqrt{\varrho^2+\eta^2}\qquad\quad R_t=
\sqrt{(1-\varrho)^2+\eta^2}  \e
The circle defined in the $(\varrho,\eta)$ plane by the value for
$B(\kpn)$ ((\ref{rhz}) and (\ref{r02})) intersects for the allowed
range of parameters with the circle
determined by a value for $R_b$.
This allows to determine $R_t$. One finds
\b\l{rt2}
R^2_t=1+R^2_b+{r^2_o-R^2_b\o\varrho_o}-\varrho_o  \e
\subsection{The Analysis of $|V_{td}|$}
The modulus of $V_{td}$ is obtained as
\b\l{vtd} |V_{td}|=A \lambda^3 R_t  \e
We note that besides $A$ (or $|V_{cb}|$) and $m_t$
the value of $R_b$ (or $|V_{ub}/ V_{cb}|$) is needed
together with the branching ratio in order to extract $|V_{td}|$ from
\kpnn decay.\hfill\break
Before discussing numerical examples we will now compile the input
parameters necessary for the analysis. We take
\b\l{mwtl}M_W=80GeV\qquad m_\tau =1.78GeV\qquad \lambda=0.22 \e
\b\l{aswb}\alpha=1/128\qquad \sin^2\Theta_W=0.23\qquad
    B(K^+\to\pi^o e^+\nu)=4.8\cdot 10^{-2}  \e
The remaining parameters are given in table 2 together with two
different sets of uncertainties. Range I corresponds to the present
status with the exception of $B(\kpn)$, where we have just chosen a
typical number. For later discussion we have also displayed range II
illustrating some possible future scenario.
\begin{table}
\begin{tabular}{|r||c|c|c|c|c|c|}\hline
&$m_t/GeV$&$m_c/GeV$&$\Lambda_{\overline{MS}}/GeV$&$A$&$R_b$&
  $B(\kpn)/10^{-10}$\\ \hline
&150&1.3&0.25&0.89&0.45&1.0\\ \hline
I&$\pm 50$&$\pm 0.1$&$\pm 0.10$&$\pm 0.08$&$\pm 0.14$&$\pm 0.5$\\ \hline
II&$\pm 5$&$\pm 0.05$&$\pm 0.05$&$\pm 0.04$&$\pm 0.05$&$\pm 0.1$\\ \hline
\end{tabular}
\centerline{}
{\bf Table 2:} Collection of input parameters with two different
choices for the uncertainties.
\end{table}
The values of $A$ and $R_b$ correspond to $|V_{cb}|=0.043\pm 0.004$
and $|V_{ub}/V_{cb}|=0.10\pm 0.03$ (range I), which are in the ball
park of various determinations.
The mass parameters $m_t$ and $m_c$
in table 2 stand for $m_t(m_t)$ and $m_c(m_c)$.\hfill\break
The various
uncertainties in all these parameters result in a rather large
range of predicted values for $B(\kpn)$, typically $(1-6)\cdot 10^{-10}$.
Similar uncertainties exist for the determination
of $|V_{td}|$ from the branching ratio \cite{BH,D}.\hfill\break
However, these uncertainties from input parameters should be
distinguished from the uncertainties arising due to the
$\mu$-dependences of the physical amplitudes, which have their origin
in the truncation of the perturbation series. The parameters could in
principle be considered as determined from somewhere else and in the
ideal case known without errors. However even in this case there would
still remain the $\mu$-scale ambiguity which constitutes the intrinsic
uncertainty of the theoretical prediction. An improvement in this issue
would require going even beyond NLLA.\hfill\break
For these reasons we will focus our attention on the scale ambiguity
and compare LLA and NLLA in this respect. As an illustration of this
point let us consider the determination of $|V_{td}|$ via (\ref{vtd}).
We fix all input parameters at their central values but allow
a variation of the renormalization scales within the ranges
$1GeV\le\mu_c\le 3GeV$ and
$100GeV\le\mu_t\le 300GeV$ in the charm- and the top-sector,
respectively. The variation of the unphysical scales $\mu_c$ and
$\mu_t$ is of course not unique. To avoid large logarithms $\mu_c$
and $\mu_t$ should however not deviate too much from $m_c$ and
$m_t$ respectively. Our ranges of $\mu_c$, $\mu_t$, chosen for
definiteness, satisfy this requirement. The ensuing variations in
predictions of physical quantities are to be understood as indications
for typical theoretical uncertainties to be expected due to the
truncation of the perturbation series.\hfill\break
We find
\b\l{vtd1}8.58\cdot 10^{-3}\le|V_{td}|\le 11.35\cdot 10^{-3}
\qquad {\rm LLA} \e
\b\l{vtd2}9.51\cdot 10^{-3}\le|V_{td}|\le 10.22\cdot 10^{-3}
\qquad {\rm NLLA} \e
The central values at $\mu_c=m_c$ and $\mu_t=m_t$ are
\b\l{vtd1c}|V_{td}|= 9.49\cdot 10^{-3}
\qquad {\rm LLA} \e
\b\l{vtd2c}|V_{td}|= 9.68\cdot 10^{-3}
\qquad {\rm NLLA} \e
One observes that including the full next-to-leading order corrections
reduces the total variation from 29\% (LLA) to 7\% (NLLA) in the
present example. The main bulk of this ambiguity stems from the charm
sector. Keeping $\mu_c=m_c$ fixed and varying only $\mu_t$, the
ambiguities would shrink to 10\% (LLA) and 1\% (NLLA). \hfill\break
These purely theoretical uncertainties should be compared to those
coming from the input parameters. To get an idea of their magnitude we
show in table 3 the relative change of $|V_{td}|$ when one parameter is
varied in the two ranges given in table 2 while all other parameters
are kept fixed. The entries in the first row indicate the varying
parameter. We note that the dependence on $\Lambda_{\overline{MS}}$
and $R_b$ is quite small.
\begin{table}
\begin{center}
\begin{tabular}{|r||c|c|c|c|c|c|}\hline
$\Delta|V_{td}|/|V_{td}|$&$m_t$&$m_c$&$\Lambda_{\overline{MS}}$&$A$&
$R_b$&$B(\kpn)$\\ \hline
I&78.5\%&11.6\%&5.7\%&16.8\%&3.4\%&70.8\%\\ \hline
II&7.2\%&5.8\%&2.8\%&8.4\%&1.2\%&13.3\%\\ \hline
\end{tabular}
\end{center}
\centerline{}
{\bf Table 3:} Relative sensitivity of $|V_{td}|$ to variations of
input parameters according to table 2.
\end{table}
It is clear that at the present stage with $B(\kpn)$ being unknown
and with a large range of possible top quark masses, the theoretical
errors related to scale ambiguities
seem rather marginal. However the determination of relevant
parameters should improve in the future. Once the corresponding
precision will have attained the level exemplified by our choice of
parameter range II in table 1, the resulting uncertainties will start
to be comparable to the theoretical uncertainty of about 7\%. Clearly,
in such a situation the gain in accuracy of the theoretical prediction
by a factor of more than 4 compared to the LLA will be very important.
In order to show this more explicitly we plot in fig.8 $|V_{td}|$
as a function of $m_t(m_t)$,
keeping all other input parameters at their central values
but varying $\mu_c$ and $\mu_t$ according to
$1GeV\leq\mu_c\leq 3GeV$ and ${2\o 3}m_t\leq\mu_t\leq 2 m_t$.
The broad and the narrow band correspond to
LLA and NLLA respectively.
This figure illustrates very clearly that the next-to-leading QCD
analysis is necessary to really exploit the clean nature of \kpnn in
future constraints on important standard model parameters.\hfill\break
To summarize: Whereas in LLA the scale ambiguity would be the main
uncertainty in $|V_{td}|$ provided $m_t$ and $B(\kpn)$ have been
accurately measured, the main uncertainty after our calculation of
next-to-leading order corrections comes from the value of
$V_{cb}$ or $A$.

\subsection{The Impact on the Unitarity Triangle}
It should be emphasized that the measurement of $B(\kpn)$ will
not only determine $|V_{td}|$ but also constrain
the shape of the unitarity triangle. Indeed the parameters $\varrho$
and $\eta$ can be determined from $B(\kpn)$ once $R_b$, $|V_{cb}|$
and $m_t$ are known. We find
\b\l{rhoe} \varrho={1\o 2}\left(\varrho_o+
         {R^2_b-r^2_o \o\varrho_o}\right)   \e
\b\l{etae} \eta=\sqrt{R^2_b-\varrho^2}  \e
with $\varrho_o$ and $r_o$ given by (\ref{rhz}) and (\ref{r02}).
$\eta$ is bound to be positive when the experimental result for
the CP violating parameter $\varepsilon$ is taken into account
\cite{BH}.\hfill\break
Here we would like to discuss the sensitivity of $r_o$, $\varrho_o$,
$\varrho$ and $\eta$ to the choice of renormalization scales. We
first note following \cite{BH} that the second term in (\ref{r02})
is negligible. Consequently
\b\l{r0a} r_o={1\o A^2 X(x_t)}\sqrt{B(\kpn)\o 4.62\cdot 10^{-11}}\e
is entirely determined by the top contribution to $\kpn$. Now as we
stressed in \cite{BB1,BB2} and in discussing $|V_{td}|$ above, the
scale dependence in $X(x_t)$ is after the inclusion of next-to-leading
order corrections reduced from $O(10\%)$ to $O(1\%)$. Consequently
the theoretical calculation of $r_o$ is fully under control once
$A$, $m_t$ and $B(\kpn)$ have been precisely measured.\hfill\break
In the absence of charm contributions one would have $\varrho_o=1$.
Thus the departure of $\varrho_o$ from unity measures the impact of
the charm contribution on the determination of the unitarity triangle.
Since in LLA there was a considerable scale ambiguity in the charm
contribution we expect this to show up also in the determination
of $\varrho_o$, $\varrho$ and $\eta$.
To summarize this discussion we have displayed in table 4 the
$\mu$-dependences of $r_o$, $\varrho_o$, $\varrho$, $\eta$,
$|V_{td}|$ and $B(\kpn)$. Again we take the central values for
all input parameters but vary $\mu_c$ and $\mu_t$ in the ranges
$1GeV\leq\mu_c\leq 3GeV$, $100GeV\leq\mu_t\leq 300GeV$.
For each quantity we show the resulting minimal and maximal value,
in NLLA as well as in LLA, together with the percentage of this
variation refering to the standard value ($\mu_c=m_c$ and
$\mu_t=m_t$). For the evaluation of $B(\kpn)$ one has to fix
$\varrho$ and $\eta$. We have chosen $\varrho=0.08$, $\eta=0.44$
which correspond to a central value of $B(\kpn)=10^{-10}$ in NLLA.
\hfill\break
\begin{table}
\begin{center}
\begin{tabular}{|r||c|c|c|c|}\hline
NLLA (LLA)&$r_o$&$\varrho_o$&$\varrho$&$\eta$\\ \hline
$\mu_{c(t)}=m_{c(t)}$&1.39 (1.37)&1.39 (1.39)&0.08 (0.10)&0.44 (0.44)
 \\ \hline
min&1.38 (1.31)&1.34 (1.24)&0.02 (-0.12)&0.44 (0.41)\\ \hline
max&1.40 (1.45)&1.41 (1.49)&0.10 (0.19)&0.45 (0.45)\\ \hline
&1.0\% (9.8\%)&5.3\% (17.6\%)& --- &2.3\% (9.7\%)\\ \hline
\end{tabular}
\end{center}

\begin{center}
\begin{tabular}{|r||c|c|}\hline
NLLA (LLA)&
$|V_{td}|/10^{-3}$&$B(\kpn)/10^{-10}$\\ \hline
$\mu_{c(t)}=m_{c(t)}$&
 9.68 (9.49)&1.00 (1.03)\\ \hline
min&
 9.51 (8.58)&0.92 (0.76)\\ \hline
max&
 10.22 (11.35)&1.03 (1.19)\\ \hline
&7.4\% (29.2\%)&10.9\% (41.0\%)\\ \hline
\end{tabular}
\end{center}
\centerline{}
{\bf Table 4:} Various quantities relevant for the phenomenology
of \kpnn and their theoretical uncertainties in LLA and in NLLA.
\end{table}
As anticipated the scale uncertainties present in LLA are considerably
reduced in NLLA. This is in particular seen in the case of $\varrho$
where the uncertainty $\Delta\varrho=0.31$ present in LLA is reduced
to $\Delta\varrho=0.08$ after the inclusion of next-to-leading order
corrections. The very small ambiguity in $\eta$ results partly from
the small value of $\varrho$ chosen in this example, but generally
$\eta$ is less sensitive to $\mu$ than $\varrho$.\hfill\break
In fig.9 we show the position of the point ($\varrho$,$\eta$) which
determines the unitarity triangle. To this end we have fixed all
parameters at their central values except for $R_b$ for which we
have chosen three representative numbers, $R_b$=0.31, 0.45, 0.59.
The full and the reduced ranges represent LLA and NLLA respectively.
The impact of our calculations on the accuracy of determining the
unitarity triangle is quite impressive. We note that the reduction
in the uncertainty in $\varrho$ may allow to resolve the two-fold
ambiguity ($\varrho >0$, $\varrho <0$) present in the analysis of the
CP-violating $\varepsilon$-parameter.\hfill\break
We note also that the 40\% uncertainty in $B(\kpn)$ present in LLA
is reduced to 11\% in NLLA. This is also seen in fig.10 where
$B(\kpn)$ is given as a function of $m_t$ for central values of the
parameters. The meaning of the curves is as in fig.8.

\section{Short Distance Contribution to \klmm Beyond Leading Logs}
\subsection{Renormalization Group Analysis}
The branching ratio $B(\klm)$ has already been measured in experiment
\cite{IM}.
However, since this decay receives important long distance
contributions, a complete theoretical treatment is more difficult
than in the case of $\kpn$. Recent discussions of this issue have been
given in \cite{GN,BG,PK}. Yet the pure short distance part of \klmm
is on the same footing as \kpnn and can be calculated by the same
methods. Since the computation is very similar in both cases, we will
restrict ourselves to mainly quote the final results for $\klm$,
indicating the most important differences to the case of $\kpn$.
\hfill\break
The effective Hamiltonian inducing the short distance contribution to
\klmm can be written as
\b\l{hklm}{\cal H}_{eff}=-{G_F \o{\sqrt 2}}{\alpha\o 2\pi \sin^2\Theta_W}
 \left( V^{\ast}_{cs}V_{cd} Y_{NL}+
V^{\ast}_{ts}V_{td} Y(x_t)\right)
 (\bar sd)_{V-A}(\bar\mu\mu)_{V-A}  \e
The resulting branching ratio is
\b\l{bklm}
B(\klm)_{SD}={\alpha^2 B(K^+\to\mu^+\nu)\o
 V^2_{us} \pi^2 \sin^4\Theta_W}{\tau(K_L)\o \tau(K^+)}
\left[ Re\ V^\ast_{cs}V_{cd} Y_{NL}+
 Re\ V^\ast_{ts}V_{td} Y(x_t)\right]^2  \e
The function $Y(x)$ is given by
\b\l{yy}
Y(x) = Y_0(x) + a Y_1(x)\e
where \cite{IL}
\b\l{yy0}
Y_0(x) = {x\over 8}\left[{4-x\over 1-x}+{3x\over (1-x)^2}\ln x\right]
\e
and \cite{BB2}
\begin{eqnarray}\l{yy1}
Y_1(x) = &&{4x + 16 x^2 + 4x^3 \over 3(1-x)^2} -
           {4x - 10x^2-x^3-x^4\over (1-x)^3} \ln x\nonumber\\
         &+&{2x - 14x^2 + x^3 - x^4\over 2(1-x)^3} \ln^2 x
           + {2x + x^3\over (1-x)^2} L_2(1-x)\nonumber\\
         &+&8x {\partial Y_0(x) \over \partial x} \ln x_\mu
\end{eqnarray}
The RG expression $Y_{NL}$ representing the charm contribution reads
\b\l{ynl} Y_{NL}=C_{NL}-B^{(-1/2)}_{NL}  \e
where $C_{NL}$ is the Z-penguin part given in (\ref{cnln}) and
$B^{(-1/2)}_{NL}$ is the box contribution in the charm sector, relevant
for the case of final state leptons with weak isospin $T_3=-1/2$.
We find
\begin{eqnarray}\l{bmnln}
\lefteqn{B^{(-1/2)}_{NL}={x(m)\o 4}K^{24\o 25}_c\left[ 3(1-K_2)\left(
 {1\o a(\mu)}+{15212\o 1875}(1-K^{-1}_c)\right)\right.}\nonumber\\
&&-\left.\ln{\mu^2\o m^2}-
  {329\o 12}+{15212\o 625}K_2+{30581\o 7500}K K_2
  \right]
\end{eqnarray}
Note the simple relation to $B^{(1/2)}_{NL}$ in (\ref{bnln}) (for
$r=0$)
\b\l{dbnl}
B^{(-1/2)}_{NL}-B^{(1/2)}_{NL}={x(m)\o 2}K^{24\o 25}_c (K K_2-1)\e
For the Z-penguin contribution the change from neutrinos to muons
as external leptons is trivial and the corresponding function $C_{NL}$
is the same in both cases. The box-part on the other hand is slightly
different for \klmm and \kpnn since the lepton line in the box
diagram is reversed. As a first consequence no internal lepton mass
appears in the case of $\klm$. The renormalization group calculation
is however very similar to the one for $\kpn$. We comment briefly on
the most important differences. Starting point are again
eqs. (\ref{hbop})--(\ref{qnub}) but with the operator $Q$ now defined as
\b\l{qbm} Q={m^2\o g^2}(\bar sd)_{V-A}(\bar\mu\mu)_{V-A} \e
Using the NDR-scheme and the projection
\b\l{projm} \gamma^\mu\gamma^\alpha\gamma^\nu(1-\gamma_5)\otimes
  \gamma_\nu\gamma_\alpha\gamma_\mu(1-\gamma_5)\to  4(1- 2
 \varepsilon) \gamma^\mu(1-\gamma_5)\otimes\gamma_\mu(1-\gamma_5)
 \e
we have instead of (\ref{mob})
\b\l{mobm}
\langle O\rangle=-a(\mu)\left( 4 \ln{\mu^2\o m^2}-2\right)
\langle Q\rangle   \e
For $B^{(-1/2)}_{NL}$ one then finds
\b\l{bmnl}
B^{(-1/2)}_{NL}={x(\mu)\o 16}\left[ -4\ln{\mu^2\o m^2}+2+
{c_2(\mu)\o a(\mu)}\right]  \e
The Wilson coefficient $c_2(\mu)$ is as given in (\ref{c2e}), however
with different values for $B_2$ and $\gamma_{12}$:
\b\l{bg12m}
B_2=2\qquad\gamma^{(0)}_{12}=8\qquad\gamma^{(1)}_{12}=160 C_F \e
Putting all this together, using (\ref{xmu}) and setting $N=3$, $f=4$
one finally obtains (\ref{bmnln}).
\subsection{Discussion of the charm contribution}
We now turn to the discussion of theoretical uncertainties. In the
top-quark sector the residual scale dependence is almost completely
eliminated when the $O(\alpha_s)$ corrections are included
(\ref{yy}), just as in the case of $\kpn$. The charm contribution is
again more problematic in this respect. In addition it exhibits a
somewhat different structure than the charm sector in $\kpn$. To see
this, recall that the function $Y$ \cite{BB2} reads in terms of the
Z-penguin- and box-contribution
\b\l{ycb} Y(x)=C(x)-B(x,-1/2)  \e
where, to linear order in $x$, $C$ is given in (\ref{cxc}) and
\b\l{bmxc} B(x,-1/2)\doteq{1\o 4}x \ln x+{1\o 4}x+a \left(
 x\ln^2 x+{23\o 3}x\ln x+{25+\pi^2\o 3}x\right)   \e
It follows that the terms of $O(x\ln x)$ and $O(a x\ln x)$ both
cancel in
\b\l{yxc} Y(x)\doteq {1\o 2}x+a\left(x\ln^2 x+{4+\pi^2\o 3}x\right)\e
We will briefly comment on the most prominent differences of $Y_{NL}$
and $X_{NL}$. First, without QCD no logarithmic terms are present in
$Y(x)$, as can be seen in (\ref{yxc}). They are to lowest order
cancelled between the Z-penguin- and the box contribution. As a
consequence the non-logarithmic term $\sim x$ is more important here
than in the case of $X_{NL}$. Logarithms are generated through QCD
and summed to next-to-leading order in the renormalization group
result (\ref{ynl}). Their effect is to enhance the zeroth order
expression by a factor of about 2.5. Recall that the function
$X(x_c)$ on the contrary is suppressed by $\sim 30\%$ through QCD
effects.
Nevertheless numerically $X_{NL}$ still exceeds $Y_{NL}$ by a
factor of four. Comparing the magnitude of the corrections in
$X_{NL}$ and $Y_{NL}$ one realizes that QCD is much more important
in $Y_{NL}$ than in $X_{NL}$.
The consequence is a larger relative $\mu$-dependence of
$Y_{NL}$. In fact one finds
$(Y_{NL}(1GeV)-Y_{NL}(3GeV))/Y_{NL}(m_c)=43.6\%$
whereas in LLA
$(\tilde{Y}_L(1GeV)-\tilde{Y}_L(3GeV))/\tilde{Y}_L(m_c)=71\%$.
Here $\tilde{Y}_L$ is defined to be the leading log function
but including the numerically important term $x(\mu)/2$,
although this is not fully consistent in this approximation.
The effect of a b-quark threshold, which we neglect, is also
larger here ($\sim 1\%$) than it was for $X_{NL}$.
\hfill\break
The $\mu$-dependences are illustrated in fig.7 where the leading log
result is shown without ($Y_L$) and including ($\tilde{Y}_L$) the
non-logarithmic term $x/2$. Since the residual $\mu$-dependence
turns out to be quite pronounced even in NLLA, one may start to worry
about the validity of perturbation theory. At least the scale
ambiguity is substantially reduced when going from leading logs to
NLLA. We will take the observed variation as an estimate for the
theoretical uncertainty even in this case and briefly discuss the
consequences in the next subsection. Fortunately the function
$Y_{NL}$ is less important for \klmm than $X_{NL}$ is
for $\kpn$. Consequently the net impact of the $\mu$-dependence on
$B(\klm)_{SD}$ will turn out to be not as dramatic as could be expected
from the discussion above.
Finally in table 5
we give numerical values of $Y_{NL}$ for various  $m_c$ and
$\Lambda_{\overline{MS}}$ with $\mu_c=m_c$.
\begin{table}
\begin{center}
\begin{tabular}{|c|c|c|c|}\hline
&\multicolumn{3}{c|}{$Y_{NL}/10^{-4}$}\\ \hline
$\Lambda_{\overline{MS}}$, $m_c$ [$GeV$]
&1.2&1.3&1.4\\ \hline
0.15&2.78&3.18&3.61\\ \hline
0.25&2.96&3.39&3.84\\ \hline
0.35&3.10&3.55&4.03\\ \hline
\end{tabular}
\end{center}
\centerline{}
{\bf Table 5:} The function $Y_{NL}$ for various $\Lambda_{\overline MS}$
and $m_c$.
\end{table}

\subsection{$B(\klm)_{SD}$ and the Parameter $\varrho$}
Expressing CKM elements through Wolfenstein parameters we find
\b\l{bklmn}
B(\klm)_{SD}=1.71\cdot 10^{-9} A^4 Y^2(x_t) [\bar\varrho_o-\varrho]^2\e
where
\b\l{qrhn}\bar\varrho_o=1+{417\o A^2}{Y_{NL}\o Y(x_t)}  \e
Formula (\ref{bklmn}) together with $Y(x_t)$ and $Y_{NL}$ in
(\ref{yy}) and (\ref{ynl}) respectively gives $B(\klm)_{SD}$
in the standard model with QCD effects in next-to-leading order.
It is the generalization of the QCD calculations in \cite{EH} where
only leading logs and some non-leading contributions have been taken
into account.
The experimental value of $B(\klm)_{SD}$ determines the value of
$\varrho$ given by
\b\l{rhr0}  \varrho=\bar\varrho_o-\bar r_o \qquad
  \bar r_o=\sqrt{B(\klm)_{SD}\o 1.71\cdot 10^{-9}}{1\o A^2 Y(x_t)}\e
Similarly to $r_o$ in the case of $\kpn$, the value of $\bar r_o$ is
fully determined by the top contribution which has only a very weak
renormalization scale ambiguity after the inclusion of
$O(\alpha_s)$ corrections. The main scale ambiguity is present in
$\bar\varrho_o$ whose departure from unity measures the relative
importance of the charm contribution. In order to illustrate the
scale ambiguities in the context of $\klm$ we show in table 6
$\bar r_o$, $\bar\varrho_o$, $\varrho$ and $B(\klm)_{SD}$
and their uncertainties. Here the same input is used as for table 4
(sect. 6.3.). To obtain the numbers in the last column we have set
$\varrho=0.09$, corresponding to $B(\klm)_{SD}=10^{-9}$ in NLLA.
\begin{table}
\begin{center}
\begin{tabular}{|r||c|c|c|c|}\hline
NLLA (LLA)&$\bar r_o$&$\bar\varrho_o$&$\varrho$&
$B(\klm)_{SD}/10^{-9}$\\ \hline
$\mu_{c(t)}=m_{c(t)}$&1.12 (1.16)&1.21 (1.20)&
 0.09 (0.04)&1.00 (0.91)\\ \hline
min&1.12 (1.10)&1.14 (1.10)&
 0.02 (-0.14)&0.88 (0.67)\\ \hline
max&1.12 (1.26)&1.24 (1.27)&
 0.12 (0.14)&1.05 (1.09)\\ \hline
&0.6\% (13.8\%)&7.6\% (13.8\%)&
 --- &16.8\% (45.8\%)\\ \hline
\end{tabular}
\end{center}
\centerline{}
{\bf Table 6:} Theoretical uncertainties in the phenomenology of
\klmm  in LLA and NLLA.
\end{table}
\hfill\break
In the case of $\varrho$ the uncertainty $\Delta\varrho=0.28$
present in LLA is reduced to $\Delta\varrho=0.10$ in NLLA. Similarly
the 46\% uncertainty in $B(\klm)_{SD}$ is reduced to 17\%. This
theoretical uncertainty in the short-distance part of \klmm left
over even after inclusion of next-to-leading order corrections
together with the poorly known long distance contributions makes this
decay unfortunately less suitable for the determination of standard
model parameters than $\kpn$.

\section{Summary}

In this paper we have given a detailed analysis of the theoretical
prediction for the rare decay \kpnn in the standard model. We have
performed the first calculation of QCD effects in this process at
the next-to-leading order. This required two different approaches,
the inclusion of the full $O(\alpha_s)$ correction to all orders in
$m_t/M_W$ for the top-quark contribution \cite{BB1,BB2}
and a two-loop renormalization
group calculation in the charm quark case performed here.
Compared to the previous leading-log estimates \cite{EH,DDG,BBH}
the main benefits of our analysis are:
\begin{itemize}
\item
The scale ambiguity in the top quark mass can be
essentially removed.
\item
The scale dependence is substantially reduced in the charm sector.
Consequently the uncertainty due to the choice of $m_c(\mu)$
stressed in \cite{D,HR} is also reduced automatically.
\item
The next-to-leading renormalization group calculation allows a
meaningful use of $\Lambda_{\overline{MS}}$ for the charm contribution.
\item
The non-logarithmic mass terms $\sim x_c$ and the effect of the
tau-lepton mass can be incorporated consistently.
\item
The renormalization scheme dependence in QCD cor\-rec\-tions to
non-lo\-ga\-rith\-mic terms addressed in \cite{DDG} can be avoided.
\end{itemize}
The consequence of these improvements is a considerably reduced
uncertainty of the theoretical prediction for $B(\kpn)$ and in the
related determination of CKM parameters from this decay.\hfill\break
As a byproduct of the analysis of \kpnn the short-distance part of
\klmm has been obtained in NLLA. In this case the scale uncertainties
are also reduced considerably compared to the LLA, but the residual
ambiguity is larger than in $\kpn$. Still $B(\klm)_{SD}$ could be
used to constrain the Wolfenstein parameter $\varrho$ with improved
accuracy in NLLA. Unfortunately however the presence of the poorly
known long distance contribution to \klmm makes this decay less
suitable for a determination of standard model parameters than
$\kpn$.\hfill\break
\kpnn could be used in particular to
constrain and eventually determine $V_{td}$. $B^o-\bar B^o$ mixing,
also sensitive to $V_{td}$, is probably less suited for this purpose
because it involves a hadronic matrix element, which is hard to calculate
reliably. By contrast, no further theoretical uncertainty is present in
\kpnn besides the residual uncertainty due to the truncation of the
(RG improved) perturbation series, which illustrates the clean nature of
this decay. For the determination of $|V_{td}|$ this ambiguity amounts to
roughly 7\%, compared to about 30\% in LLA, for typical parameter values.
The remaining uncertainty is almost entirely due to the charm sector.
\hfill\break
We have also analyzed the impact of next-to-leading corrections to
\kpnn on the determination of the parameters ($\varrho$, $\eta$),
relevant for the unitarity triangle. The inclusion of these
corrections considerably reduces the uncertainties
in ($\varrho$, $\eta$) present in LLA.\hfill\break
We remark in this context that, since the charm sector does not matter
for the CP violating mode $K_L\to\pi^o\nu\bar\nu$, the standard model
prediction in this case is extremly accurate after the inclusion of QCD
corrections. In fact this decay is then essentially free of theoretical
uncertainties, which makes the challenging search for
$K_L\to\pi^o\nu\bar\nu$ particularly desirable.
Similar comments apply to $B\to X_s\nu\bar\nu$. For details see
\cite{BB2}.\hfill\break
The residual theoretical ambiguity in \kpnn is much smaller than
uncertainties coming from the various input parameters involved.
Clearly, to be able to take advantage of the clean theoretical
prediction, further progress has to be made on this side. The most
important tasks for the future are the determination of the
top quark mass, a more accurate determination of $|V_{cb}|$ and
above all -- hopefully -- a measurement of $\kpn$.

\vfill\eject

\section*{Figure Captions}

\newcounter{fig}
\begin{list}{\bf Fig. \arabic{fig}}{\usecounter{fig}\labelsep0.4cm
 \labelwidth1.6cm}

\item The leading one-loop Z-penguin- and box diagrams inducing
$\bar sd\to\bar\nu\nu$, responsible for $\kpn$.
\item The Z-penguin in the effective theory relevant for the charm
contribution (W-, Z-boson integrated out). The square blob denotes
operator insertion.
\item Gluonic corrections to the diagrams of fig.2. Symmetric
diagrams and counterterm contributions
are not shown explicitly. The last diagram vanishes for massless,
on-shell external quarks in dimensional regularization.
\item The box diagram in the effective theory.
\item Gluonic corrections to the diagram of fig.4.
\item The function $X$ representing the charm contribution to the
$\bar sd\to\bar\nu\nu$ amplitude (for $m_l=0$) and its dependence on the
renormalization scale $\mu_c$, reflecting the theoretical uncertainty.
The cases shown are: $X_L$ (LLA), $X_{NL}$ (NLLA) and $X_0$
(without QCD).
\item The scale dependence of the function $Y$, relevant for
$\bar sd\to\bar\mu\mu$. We show the results in LLA, with
($\tilde{Y}_L$) and without ($Y_L$) non-logarithmic mass term,
in NLLA ($Y_{NL}$) and without QCD.
\item $|V_{td}|$ determined from $B(\kpn)$ as a function of
$m_t=m_t(m_t)$ for fixed values of the necessary input parameters
(see sect. 6). The narrow (broad) band results from a variation of
the renormalization scales $\mu_c$, $\mu_t$ in the ranges
$1GeV\leq\mu_c\leq 3GeV$ and ${2\o 3}m_t\leq\mu_t\leq 2m_t$ in
NLLA (LLA) and indicates the theoretical uncertainty involved in
this analysis.
\item Theoretical uncertainties in the determination of the
unitarity triangle in the ($\varrho$,$\eta$) plane from $B(\kpn)$.
For fixed parameter values the vertex of the unitarity triangle is
forced to lie on a circle of radius $R_b$ around the origin.
The variation of the scales $\mu_c$, $\mu_t$ within
$1GeV\leq\mu_c\leq 3GeV$, $100GeV\leq\mu_t\leq 300GeV$ then yields
the indicated ranges in LLA (full) and NLLA (reduced). We
show the cases $R_b$ = 0.31, 0.45, 0.59.
\item The scale ambiguity in $B(\kpn)$, in analogy to fig.8.

\end{list}
\end{document}